\documentstyle[manuscript,aps]{revtex}
\begin{document}
\draft
\title{The Sol-Gel Process Simulated by Cluster-Cluster Aggregation}

\author{Anwar Hasmy and R\'emi Jullien}
\address{Laboratoire de Science des Mat\'eriaux Vitreux, UA 1119 CNRS,
      Universit\'e Montpellier II, Place Eug\`ene Bataillon,
               34095 Montpellier Cedex 5, France}

\date{\today}
\maketitle

\begin{abstract}
The pair-correlation function $g(r,t)$ and its Fourier transform, the structure
factor $S(q,t)$, are computed during the gelation process of identical
spherical particles
using the diffusion-limited cluster-cluster aggregation model in a box. This
numerical analysis shows that the time evolution of the characteristic
cluster size  $\xi$
exhibits a crossover close to the gel time $t_g$
which depends on the volumic fraction $c$. In this model $t_g$
tends to infinity when the box size $L$ tends to infinity.
For systems of finite size, it is shown numerically
that,  when $t<t_g$,
the wave vector $q_m$, at which
$S(q,t)$ has a maximum, decreases as $S(q_m,t)^{-1/D}$,
where $D$ is an apparent fractal dimension of clusters, as measured from the
slope of
$S(q,t)$ . The time evolution of the mean number of particles per cluster
$\bar{n}$
is also investigated. Our numerical results
are in qualitative agreement with small angle scattering experiments
in several systems.
\end{abstract}

\eject

\section {INTRODUCTION}

In the last years, the kinetics of aggregation has been widely studied
\cite{fl}.
It has been studied experimentally in a large number
of systems like aqueous metal colloids \cite{whl},
silicon tetramethoxyde \cite{mk} and tetraethoxyde \cite{cdd},
alumino-silcates \cite{pbd}, colloidal silica \cite{fgp},
polystyrene \cite{bc,cg,re,hbh},
and oil in water emulsions \cite{bmg1,bmg2}.
Using small-angle scattering
techniques it has been suggested that some of these systems
\cite{pbd,bc,cg,re,bmg1,bmg2}
grow like diffusion-limited cluster-cluster aggregation (DLCA)
\cite{me,kbj,jb}.
Numerically, the gelation process in DLCA has been investigated
by analyzing   the cluster size distribution and the mean cluster size, as a
function of the time $t$ using both the Smoluchowski equation \cite{ts}
and the DLCA model in 2
dimensions \cite{kh}. On the other hand, recently \cite{haf,hfa}
it has been shown
that, when the volumic fraction $c$, or concentration, is greater than a
characteristic gel concentration $c_g$, DLCA
model leads to a homogeneous gelling network of connected
fractal clusters of mean
size $\xi$.

In some systems \cite{cdd,pbd,cg,re,hbh,bmg1,bmg2} where
fractal clusters grow separately
until obtaining a system of connected fractal clusters at large $t$,
the wave-vector-dependent scattering function $S(q,t)$ exhibits
a maximum in the wave vector $q_m$-value (which is inversely proportional
to $\xi$) at a given time $t$. This maximum can be interpreted as a crossover
between the fractal regime (intermediate $q$-values where
$S(q,t) \propto q^{-D}$) and the homogeneous regime (small $q$-values).
Moreover, some very interesting relations on the $S(q)$ curve
(see below equations 7)
have been established on various systems \cite{cg,hbh}. Since
scaling relation are known to be of wide
applications (``universality'') it is worth studying  them here.
Another interesting experimental results \cite{pbd} reveal that the gyration
radius $R_g$ (which is proportional to $\xi$) saturates close to $t_g$,
the time where the system can be considered as a gelling network.

In this paper, we present some new results obtained from
our numerical study of the gelation process using diffusion
limited cluster-cluster aggregation in a box \cite{me,kbj,jb},
where a sufficiently large
initial concentration has been considered in order to obtain a gelling
network of connected clusters at the end of the aggregation process.
We have calculated
numerically the mean number of particles per cluster $\bar{n}$
and their mean size $\xi$ as a function of time $t$ as well as
the correlation function $g(r,t)$, and the scattering function $S(q,t)$.
We show that there exists  a crossover at $t_g$ such that, for $t<t_g$,
$\xi$ increases when $t$
increases, and for $t>t_g$, the characteristic length $\xi$ remains constant
and equal to its value obtained near $t_g$.
The location of this crossover is pushed toward
the low-$t$ values when $c$ increases. For $c \gg c_g$
the normalized gel time $t_g/t_{tot}$ does not change. Additionally,
the maximum of the $S(q,t)$ curve, $S(q_m,t)$, increases as $\xi^{D_{sl}}$
where $D_{sl}$ is an apparent fractal dimension.
We report on the discrepancies between the actual fractal dimension $D$
and $D_{sl}$ for large $c$-values.
Finally, we discuss the qualitative agreement between our simulations and
some experiments.

\section{Constraints on Theory}

We have modelized the gelation process considering a three dimensional
off-lattice extension of the original diffusion cluster-cluster
aggregation model \cite{me,kbj} as in ref. \cite{hfa}
where we have shown that such a model
is well adapted to describe the aerogel structure.
During the aggregation process, we can naturally define a ``Monte-Carlo''
time $t_{mc}$ which is increased by an arbitrary constant amount $\delta
t_{mc}$
at each iteration. Due to the use of formula (3) of ref. \cite{hfa}
to choose the aggregates, the actual
``physical'' time $t$ should be increased by the amount $\delta t$ related to
$\delta t_{mc}$ by:
\begin{eqnarray}
\delta{t}={\delta t_{mc} \over {\sum_i{n_i^\alpha}}}
\label{E3}
\end{eqnarray}
In our calculation the time unit has been arbitrarily fixed by setting $\delta
t_{mc}=1$.

Using scaling arguments, it can be established that the whole system remains
invariant
if length, mass and time are conveniently rescaled
simultaneously \cite{bj,k}. Changing the average mass $\bar{n}$ by a factor
$b$, the distance should be rescaled by $b^{1/D}$ and,
using a general mean-field argument
to evaluate the average time it takes for all clusters to pair up, the time
should
be rescaled
by $b^{1-\alpha - {(d-2)\over D}}$, giving:
\begin{mathletters}
\begin{eqnarray}
n     \propto t^\gamma
\label{E4}
\end{eqnarray}
with
\begin{eqnarray}
\gamma={1 \over {1- \alpha -{d-2 \over D}}}
\label{E5}
\end{eqnarray}
\end{mathletters}
For $\alpha = -1/D$ and $d=3$ this reasoning gives $\gamma=1$.
In figure 1 we show the time evolution  of the mean number
of particles per cluster $\bar{n}$
for three different concentrations. When $c < c_g$
$\bar{n}$ is roughly proportional to $t$ as expected from equation (2).
However, for $c > c_g$ this dependence is only observed
at short times (for $t \ll t_g$). At large times (for $t \gg t_g$)
if one still assumes
$\bar{n} \sim t^\gamma$, one should consider a very  large exponent
$\gamma \sim 4$. Note that in refs. \cite{bj,k}, equation (2)
was obtained under the    hypothesis that  the entire system remains space and
time scaling
invariant.
This hypothesis is no longer valid when a gel is formed since the system
becomes homogeneous
for distances larger than
 $\xi$ \cite{haf,hfa} and if one tries to define its global
fractal dimension it should be taken as the
dimension of space. Qualitatively, the observed increase of $\gamma$ can be
understood as
a consequence of the increase of the fractal dimension. An earlier work
by Gonzalez \cite{g1} suggests
the same conclusion. In his paper it is shown that when
$D$ increases from 1.87 to 2.05, $\gamma$ increases from 1.3 to 2.5 when using
a cluster-cluster aggregation model with a sticking probability $p$
varying from 0.5 to 0.005, reaching the chemically-limited
cluster-cluster aggregation (CLCA) model \cite{jk}, also called
reaction-limited cluster-cluster aggregation (RLCA) \cite{bb}.
An increasing $\gamma$-value with $D$ was also found in another  CLCA model
also recently proposed by Gonzalez \cite{g2}.

The distance  that a cluster must travel before colliding
with another cluster decreases when  $c$ increases so that both the gel time
$t_g$,
when it exists (i. e. for $c>c_g$), and the time
at the end of the aggregation process $t_{tot}$ should decrease as $c$
increases.
This is well observed in figure 2 where $t_g$ and $t_{tot}$ have been plotted
as a function of $c$ in a log-log plot. For  $c<c_g$
we observe that $t_{tot}$ is roughly inversely proportional to $c$ while, for
$c>c_g$, both $t_g$ and $t_{tot}$ decrease roughly as $c^{-1.75}$
($c_g$ is determined as explained in ref. \cite{hfa}).
The latter result can be simply understood if one assumes that when $c>c_g$ the
scaling
relation (2) holds up to $t\sim t_g$ where the system
is made of clusters of  mean size $\xi$  containing $\xi^D$ particles.
The concentration being related to $\xi$ by $c\sim \xi^{-(3-D)}$ (see ref.
\cite{hfa}), the relation (2)
with $\gamma = 1$ gives:
\begin{eqnarray}
t_g \propto c^{-{D \over 3-D}}
\label{E6}
\end{eqnarray}

The observed exponent of -1.75 corresponds to $D\sim 1.9$ a value larger but
quite close
to the expected value of the fractal dimension of DLCA clusters.
We point out that in our model
$c_g$ decreases as  $L^{-(3-D)}$ as showed elsewhere \cite{hfa}, and
from equation (3) one can deduce that $ t_g \rightarrow \infty$ when $L
\rightarrow \infty$. From figure 2 we also find that the ratio $t_g/t_{tot}$ is
almost
independent on $c$, for $c \gg c_g$. This implies that the time taken by
smaller clusters to stick to
the gelling network remains a finite fraction of the total time.

The two-point correlation function $g(r,t)$ can be calculated by
choosing a given path $\delta r$, and calculating the number $\delta n$ of
interparticle distances lying between $r$ and $r+\delta r$,
taking care of the periodic boundary conditions when
investigating regions outside of the box. In order to
normalize $g(r,t)$ to unity when r tends to infinity, $g(r,t)$ was
defined as follows:

\begin{eqnarray}
g(r,t)={\pi \over 6c}{1 \over 4 \pi r^2} {\delta n \over {\delta r}} ={1 \over
24cr^2}{
\delta n \over {\delta r}}
\label{E7}
\end{eqnarray}

The static structure factor, or scattering function, $S(q,t)$ in a system
containing identical particles with mean volume fraction $c$ is
given by \cite{fs}:
\begin{mathletters}
\begin{eqnarray}
S(q,t)=1+{6c \over \pi} \int^{r_m}_0{(g(r,t)-g_0){\sin qr \over qr}{4\pi r^2
dr}}
\label{E8}
\end{eqnarray}
In theory one should take $g_0=1$ and $r_m=\infty$. But, as explained in detail
in refs. \cite{haf,hfa}
we have calculated $g_0$ from the formula:
\begin{eqnarray}
g_0={\int_0^{r_m}g(r,t)4\pi r^2 dr +\pi/6c\over \int_0^{r_m}4\pi r^2 dr}
\label{E9}
\end{eqnarray}
\end{mathletters}
which gives a $g_0$ value very close to one that insures that
$S(q,t)\rightarrow 0$ when $q
\rightarrow 0$.
On the other hand, $r_m$
was chosen
equal to $L\over 2$, to avoid boundary
artifacts due to the periodic boundary conditions considered in our problem.

\section{Results}

Using equation (4), we have calculated  $g(r,t)$ at all times $t$ during the
aggregation process;
we have observed  that all the $g(r,t)$ curves  exhibit a   delta peak at
$r=1$, a   discontinuity
at $r=2$ and a   minimum at $r=\xi$, that correspond
to the typical short and long range
features of DLCA aggregates \cite{hfa}. In figure 3 we have plotted
several $g(r,t)$
curves for $c=0.005$ at four different $t$ emphasizing
the region near the minimum in order to show
that its location
(which defines the characteristic size $\xi$ of the aggregates) is
shifted to  higher $r$-values when $t$ increases.
The function $g(r,t)$ was averaged up to 20 independent simulations, and the
box size $L$ was taken equal to 103.

Figure 4 shows $\xi$ as function of $t/t_{tot}$ for four different values of
$c$. From equation (2), one expects that $\xi$ should increase as
${(t/t_{tot})}^{1 \over D}$ for $t<t_g$. However, for concentrations
close to $c_g$ this power law is only verified for $t\ll t_g$.
For $t$ smaller than but close to $t_g$ one observes a crossover regime with a
smaller slope
which extends down to smaller times when $c$ is smaller. This smaller slope
might be due to
the interpenetration of clusters which becomes important as $t$ approaches
$t_g$. The cluster
size $\xi$ saturates for $t>t_g$ suggesting that, in this regime, the remaining
free small clusters
diffuse towards the gelling network on which they stick without changing the
mean size of the
large connected clusters. However, during this regime, the density of clusters
increases and
therefore their fractal dimension should become larger. This can be checked in
figure 5
where we have plotted the fractal dimension $D$ as a function of $c$. The upper
curve
(open squares) correspond to the ''true'' fractal dimension, i.e. the
 one estimated in real space
from the power law
behavior
$g(r,t) \propto r^{D-3}$, for $3<r<\xi$.

In figure 6 we show typical $S(q,t)$ curves at different times.
These  curves exhibit the three regimes explained elsewhere
\cite{hfa},  damped oscillations at large  $q$-values $(q \gg a^{-1})$,
a   power law $S(q,t) \propto q^{-D}$ at the intermediate $q$-values, and a
vanishing
regime
for $q \ll \xi^{-1}$.
In this figure we clearly see that $q_m$ is shifted to  lower $q$-values
as   $t$ increases. This is in agreement with the results of figure 4 if one
considers that $q_m$ is
proportional to $\xi^{-1}$ (as found for $t=t_{tot}$ in ref. \cite{hfa}).
Additionally, it has been shown \cite{f} that
the scaled scattering function $S(q/q_m,t)$ and the position
of the peak $q_m$ are related by:
\begin{eqnarray}
S(q/q_m,t)=q_m(t)^{-\delta} F(q/q_m)
\label{E10}
\end{eqnarray}
where $F(q/q_m)$ is a time-independent scaling function.
In spinodal decomposition $\delta = d=3$, while in other growth problems
$\delta =D$ \cite{cg,afl}. In our case,  as showed in
inset in figure 6, the best collapse   is obtained with  $\delta=1.9$ which
roughly
corresponds to our previous estimate of the fractal dimension.
Also, this scaling behavior can be deduced by considering that
the maximum of the scattering function $S(q_m,t)$ is roughly
proportional to the mean number of particles per cluster \cite{hfa}:
\begin{mathletters}
\begin{eqnarray}
S(q_m,t) \propto \xi^{D}
\label{E11}
\end{eqnarray}
implying:
\begin{eqnarray}
q_m \propto {1\over \xi} \propto S(q_m,t)^{-{1 \over D}}
\label{E12}
\end{eqnarray}
\end{mathletters}

In figure 7 we show a log-log plot of $q_m (\approx {\pi \over \xi})$
versus $S(q_m,t)$.  For convenience we have
reported the results for $q_m={\pi \over \xi}$ by estimating $\xi$ from the
minimum of $g(r,t)$
since in such estimation the errors are smaller than estimating $q_m$
directly from $S(q,t)$.
As expected from equation (7b) a quasi linear behavior is found for the four
concentrations reported in the figure. By performing a linear regression we
have estimated the
exponents of (7b) which have been reported in table I together with the
related fractal dimension that we call $D'$. In this table we have also
reported
the values $D_{sl}$  obtained from the slopes of the respective $S(q,t)$ curves
at the last stage of the aggregation process as well as the actual fractal
dimension
$D$ obtained from
$g(r,t)$. It is shown that, while $D$ increases,  $D'$ and $D_{sl}$
decrease when  $c$ increases,
as already seen on figure 5. Similar discrepancies between
$D$ and $D_{sl}$ were observed
by Amar et al. \cite{afl} in a sub-monolayer molecular beam epitaxy model
in which the aggregation regime was described by a system of
connected fractal clusters. These        discrepancies between
$D$ and $D_{sl}$ might be explained by the fact that
for $D>2$ the single scattering theory considered  here is not
completely justified\cite{b}. Note that, for $c>c_g$ one has $D=3$ for
distances larger than  $\xi$,
while for $c \leq c_g$ the final system is a single fractal aggregate,
with $D_{sl}=D$ as showed elsewhere \cite{vpd}.

\section{Discussion}

In this section we would like to discuss our results in comparison with
several experiments.
Small-angle neutron-scattering (SANS) experiments in the sol-gel
process of alumino-silicates have been performed by Pouxviel et al.
\cite{pbd} showing that the radius of gyration $R_g$ saturates when $t$
approaches
$t_g$ (see figure 2 of ref. \cite{pbd}).
These authors have considered three different composition samples prepared
under
basic conditions and they obtained a fractal dimension
$D \approx 1.8$ as in the DLCA case. Therefore, their results
are in agreement with our numerical simulations.
Another interesting study has been reported in
ref. \cite{cg}, where small-angle light-scattering experiments have been
performed on
polystyrene spheres diluted in a water-heavy-water mixture; it has been
shown that $I(q_m)$ scale as ${q_m}^z$, where $z$ is equal to -0.58
(close to the exponent  reported in table I), at
early stage of the growth process, and at larger $t$ values it has been found
that $z$ becomes close to (-1/3)  the typical value for
spinodal decomposition.
They conclude that at early stage of spinodal decomposition
the growth is dominated by DLCA.
Similar kind of analysis
has been recently done by Hobbie et al.\cite{hbh} concerning SANS
experiments in
a hydrogen-bonded polymer blend. But in their experiments they found $D=2.4$ a
value consistent
with diffusion limited particle-cluster aggregation \cite{ws}.

Other interesting results have been reported by Bibette et al.
\cite{bmg1,bmg2}.
They have performed small angle light scattering experiments
in  monodisperse droplets
of water emulsions in oil. They have suggested that the emulsion growth is
like diffusion-limited cluster-cluster aggregation, obtaining
$D \approx 1.8$. We have extracted the $I(q_m,t)$ points from their
curves of figure 2 in ref. \cite{bmg1}, that we have compared with the
points obtained from our simulation when using the same $c$ value as
in the experiments ($c=0.005$). In figure 8 we show the good
agreement with the experiments, obtaining the same increasing behavior
of $I(q_m,t)$ as a function of $t$.

\section{conclusion}

In conclusion, in this work we have shown that numerical simulations
of DLCA model in a box can  account for the growth process
of several experimental systems.
On the other hand, our numerical calculation of $S(q,t)$ during the
DLCA process is an original investigation that extends previous numerical
analysis of this growth process \cite{fl,ts,kh,bj}, leading to the theoretical
explanation of additional interesting properties during the aggregation process
and
confirming some predictions made in
some experimental studies. In particular the power law given in equation (7b)
and the shape of our $S(q,t)$ curves are experimentally observed. Similar
calculations with the CLCA model are in progress in order to interpret some
experiments which are more likely explained by such kind of growth process.

One of
us (A. H.) would like to acknowledge support from CONICIT (Venezuela).

\begin{table}
\begin{tabular}{ccccr}
$c$& exponent $({-1 \over D'})$&$D'$&$D_{sl}$&$D$\\
\tableline
0.005& -0.52 & 1.92 & 1.79 & 1.95 \\
0.010& -0.54 & 1.85 & 1.77 & 2.02 \\
0.015& -0.55  & 1.82 & 1.76 & 2.11 \\
0.030& -0.62 & 1.61 & 1.62 & 2.32 \\
\end{tabular}
\bigskip
\caption{For four concentrations considered in the simulations, we have
reported the respective exponents ($-1 \over D'$) from equation (7b) and the
related fractal dimensions $D'$, which is compared with the apparent fractal
dimension $D_{sl}$ measured from the slope of the $S(q,t)$
curve and the ``true'' fractal dimension $D$ as deduced from $g(r,t)$.}

\end{table}


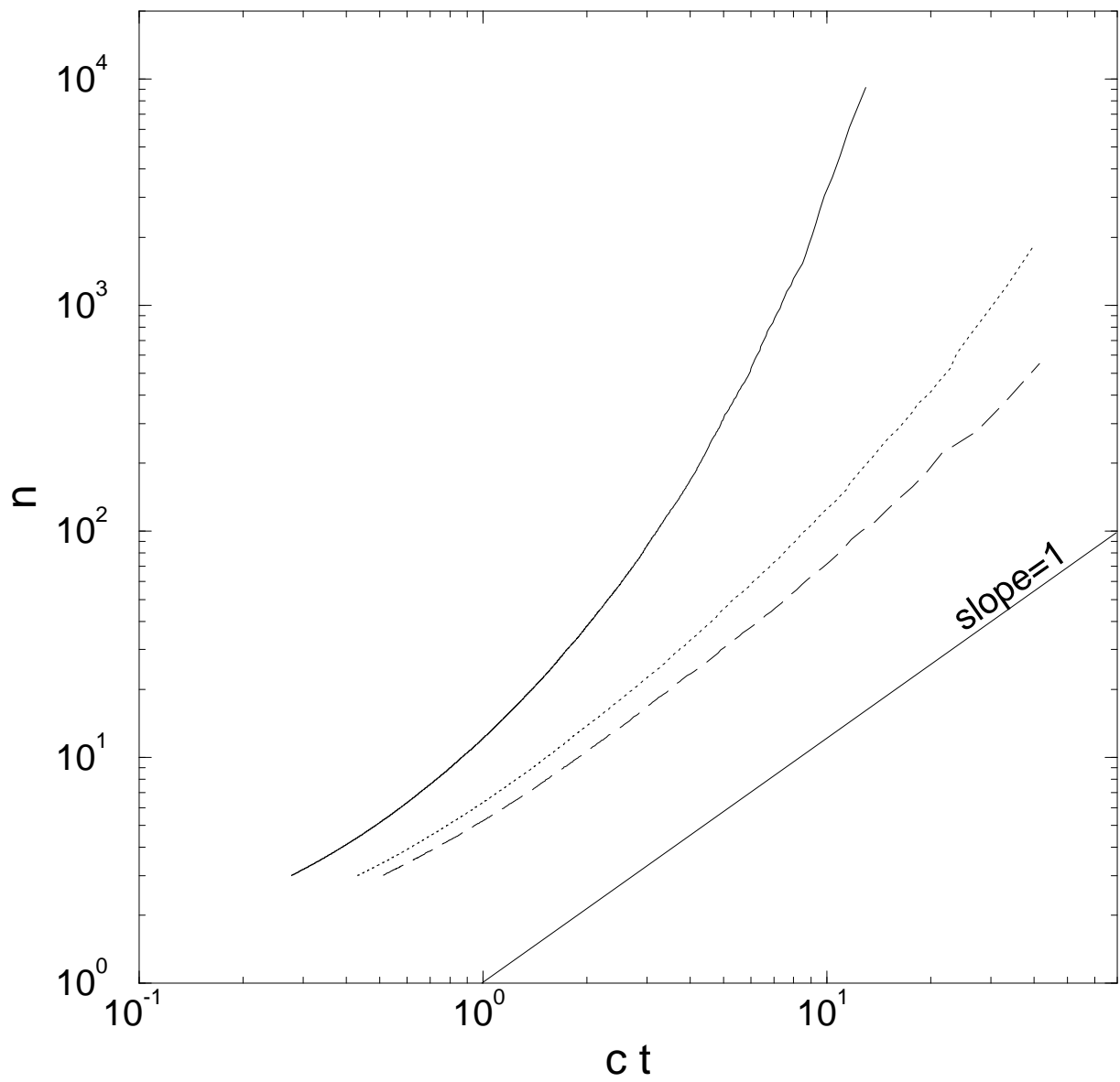
\begin{figure}
\caption{Log-log plot of $\bar{n}$ versus $c$$t$ for L=57.7, and for
$c=0.003$ (dashed line), $c=0.01$ (dotted line) and $c=0.05$ (solid line).
These curves result from averages over 20 simulations.}
\end{figure}

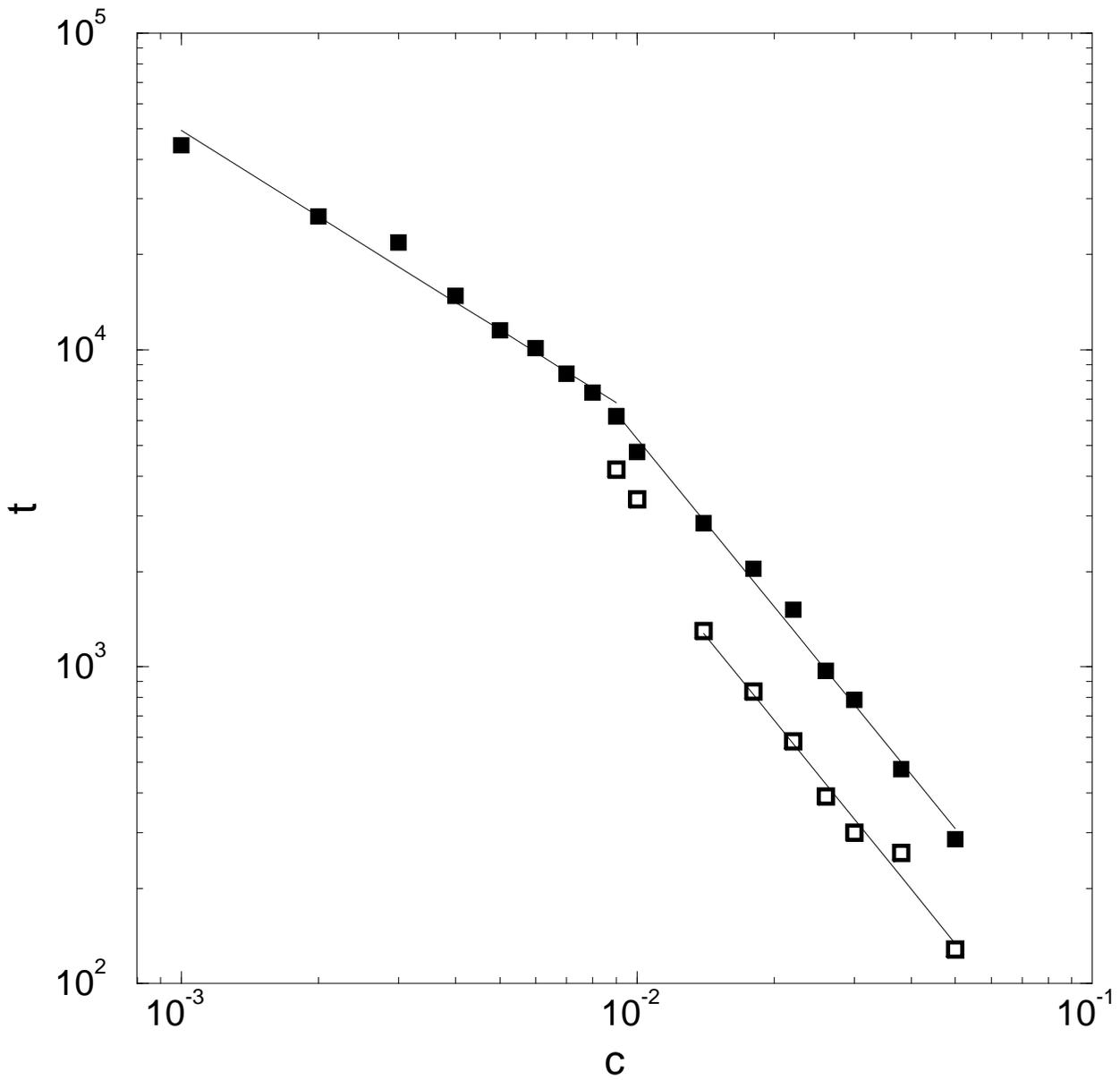
\begin{figure}
\caption{Log-log plot of $t_{tot}$ (black symbols), and $t_g$ (open symbols)
versus $c$ for L=57.7. The solid line corresponds to linear regressions and
the vertical arrow show the location of $c_g$. These data result from averages
over 20 simulations.}
\end{figure}

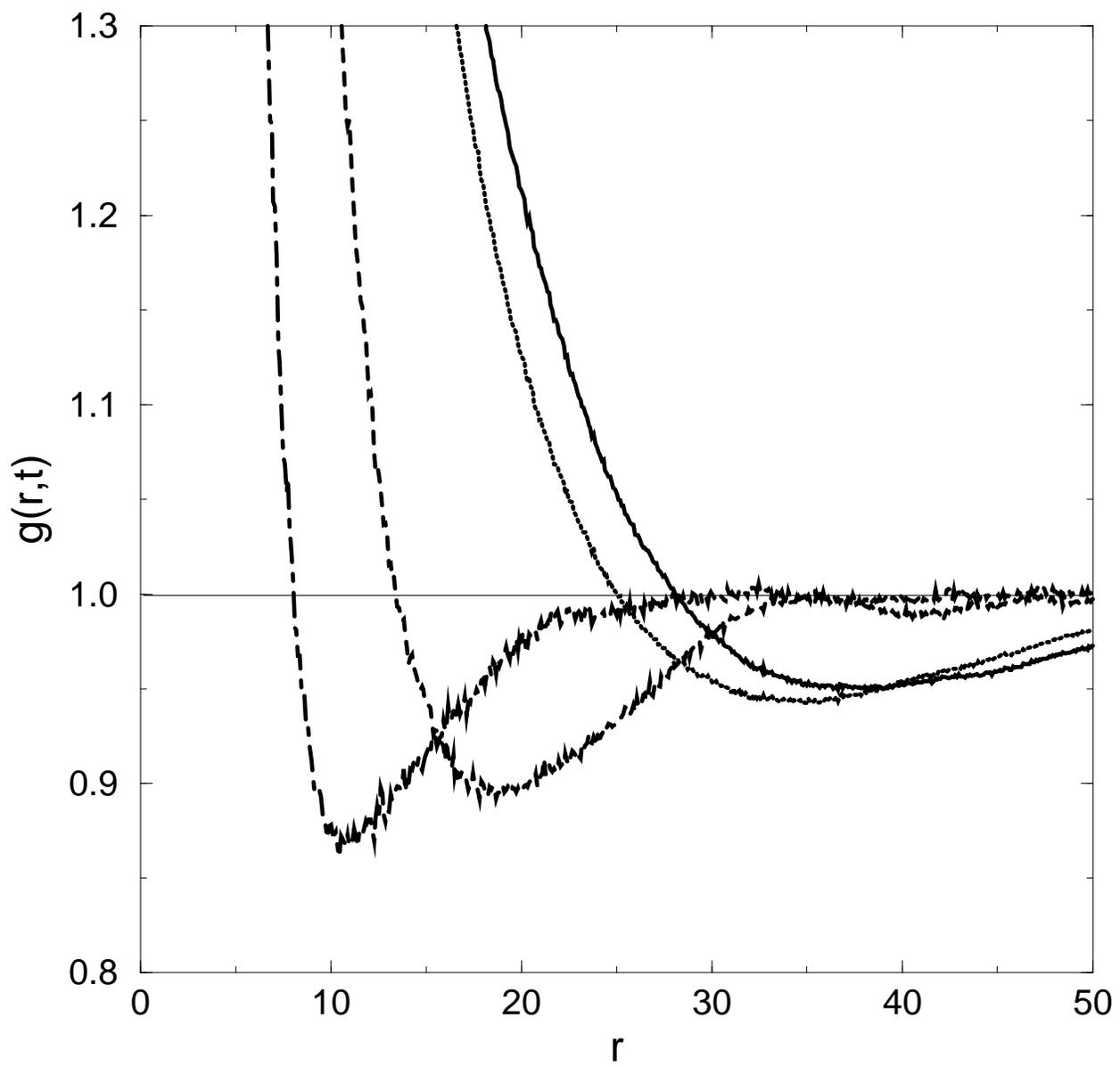
\begin{figure}
\caption{Plot of $g(r,t)$ versus $r$ for $L=103$ and $c=0.005$ at different
times $t$
of the aggregation process, $t/t_{tot}$=0.027 (dot-dashed line),
$t/t_{tot}$=0.01 (dashed line), $t/t_{tot}$=0.57 (dotted line) and
$t/t_{tot}$=1 (solid line). These curves result from averages over 10
simulations.}
\end{figure}

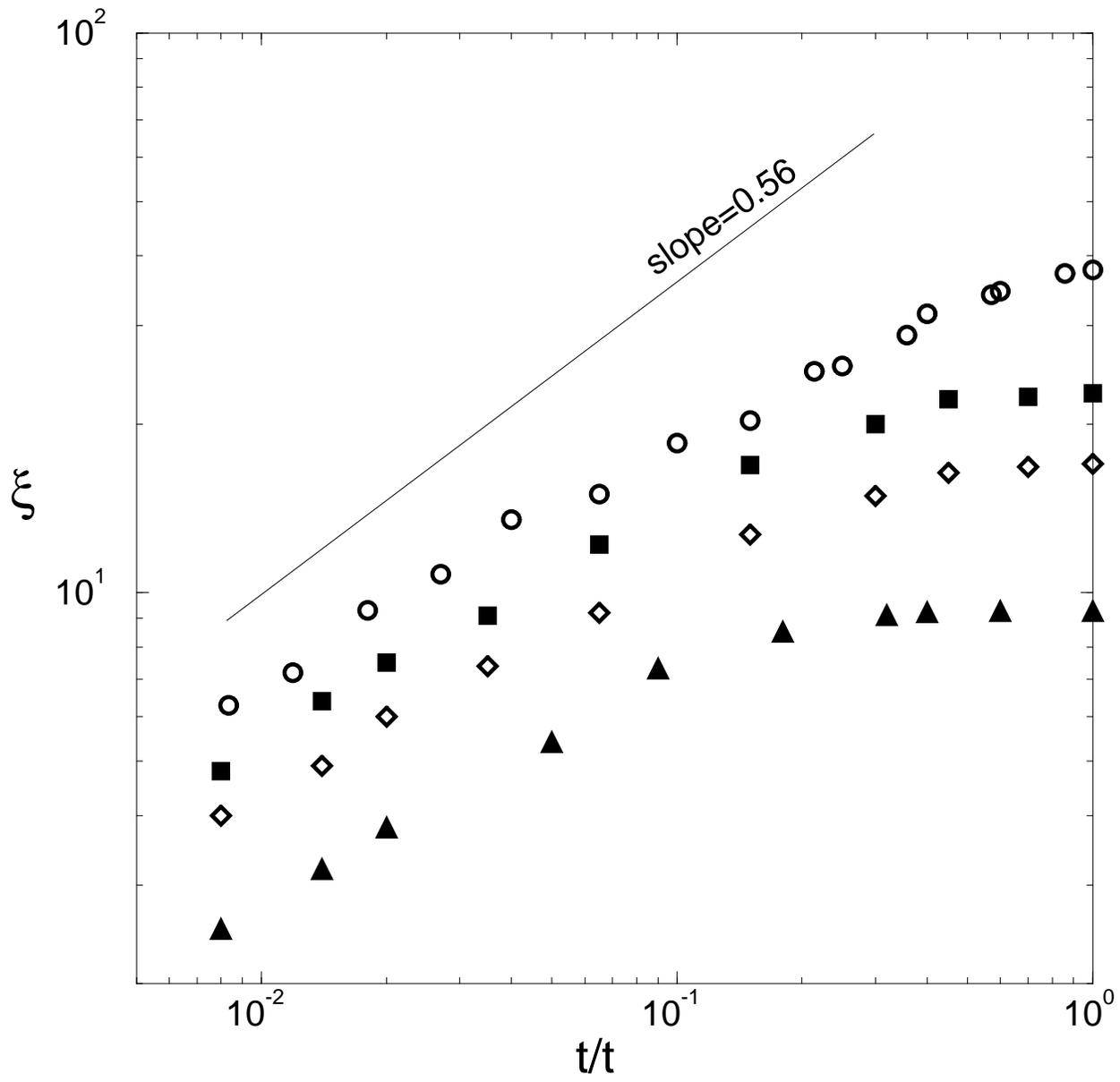
\begin{figure}
\caption{Log-log plot of $\xi$ versus $t/t_{tot}$ for $L=103$, and for
$c=0.005$ (open circles), $c=0.01$ (black squares), $c=0.015$ (open diamonds)
and $c=0.03$ (black triangles). The vertical arrows indicate the respective
location of $t_g$.
These data result from averages over 10
simulations.}
\end{figure}

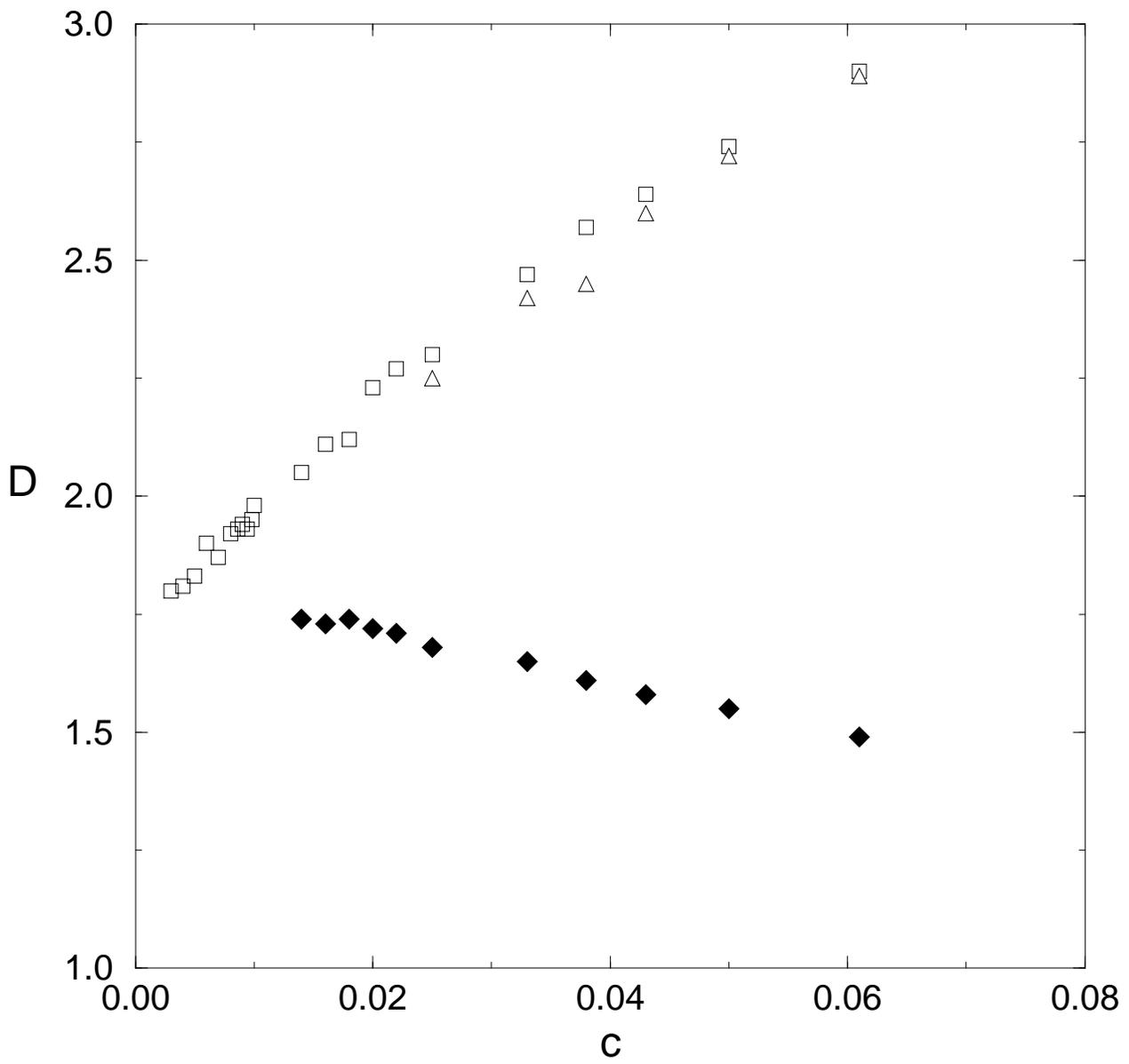
\begin{figure}
\caption{Log-log plot of the ``true'' fractal dimension $D$ (open symbols)
versus $c$, the open squares and open triangles correspond to $L=57.7$ and
$L=28.85$, respectively. The black symbols denote the $D_{sl}$ measured
from the slope of the $S(q,t)$ curve for $L=57.7$. These curves result from
averages over 20 simulations.}
\end{figure}

\begin{figure}
\caption{For four different times  $t/t_{tot}$=0.027 (dot-dashed line),
$t/t_{tot}$=0.01 (dashed line), $t/t_{tot}$=0.57 (dotted line) and
$t/t_{tot}$=1 (solid line) we show the Log-log plot of $S(q,t)$ versus $q$.
Inset: Log-log plot
of $S(q/q_m,t)q_{m}^D$. The parameters are the same as in figure 3.}
\end{figure}

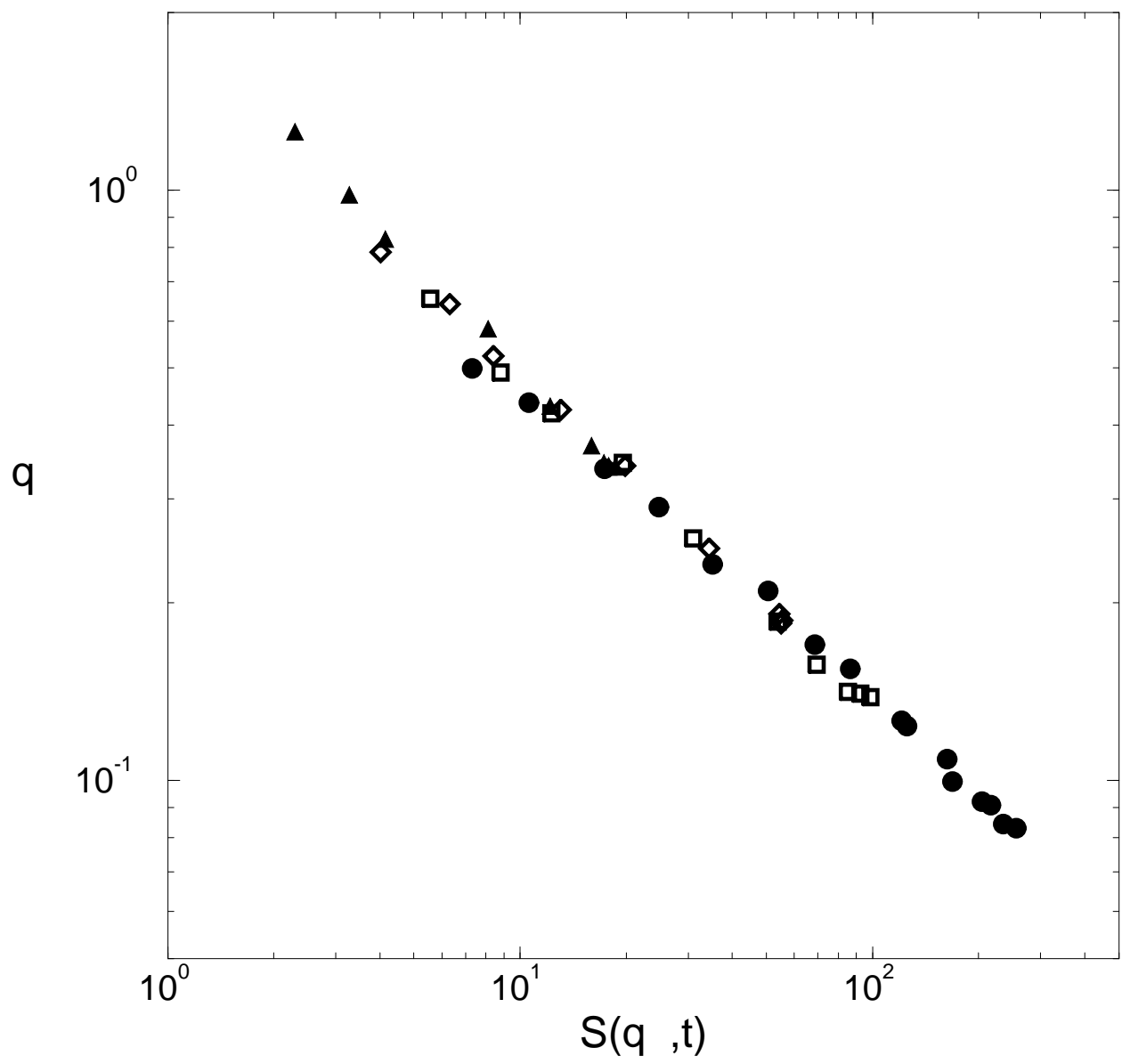
\begin{figure}
\caption{Log-log plot of $q_m$ versus $S(q_m,t)$ for $L=103$, and
for $c=0.005$ (black circles), $c=0.01$ (open squares), $c=0.015$
(open diamonds) and $c=0.03$ (black triangles). These  curves result from
averages over 10 simulations.}
\end{figure}

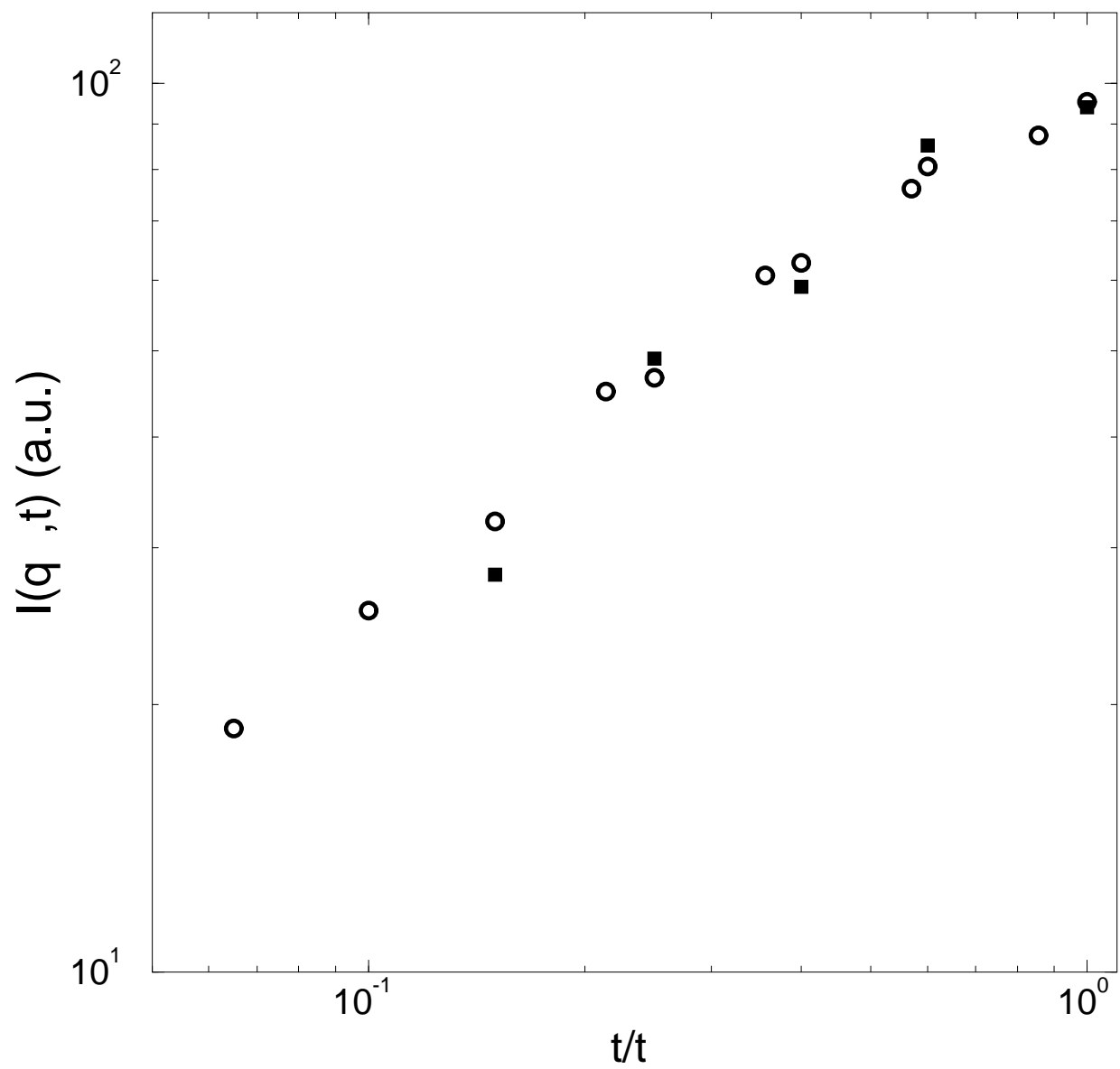
\begin{figure}
\caption{Log-log plot $I(q_m,t)$ versus $t_{tot}$. The open symbols denote
the simulation (same paprameter as in figure 3), and black symbols
denote experimental data from ref. [11].}
\end{figure}

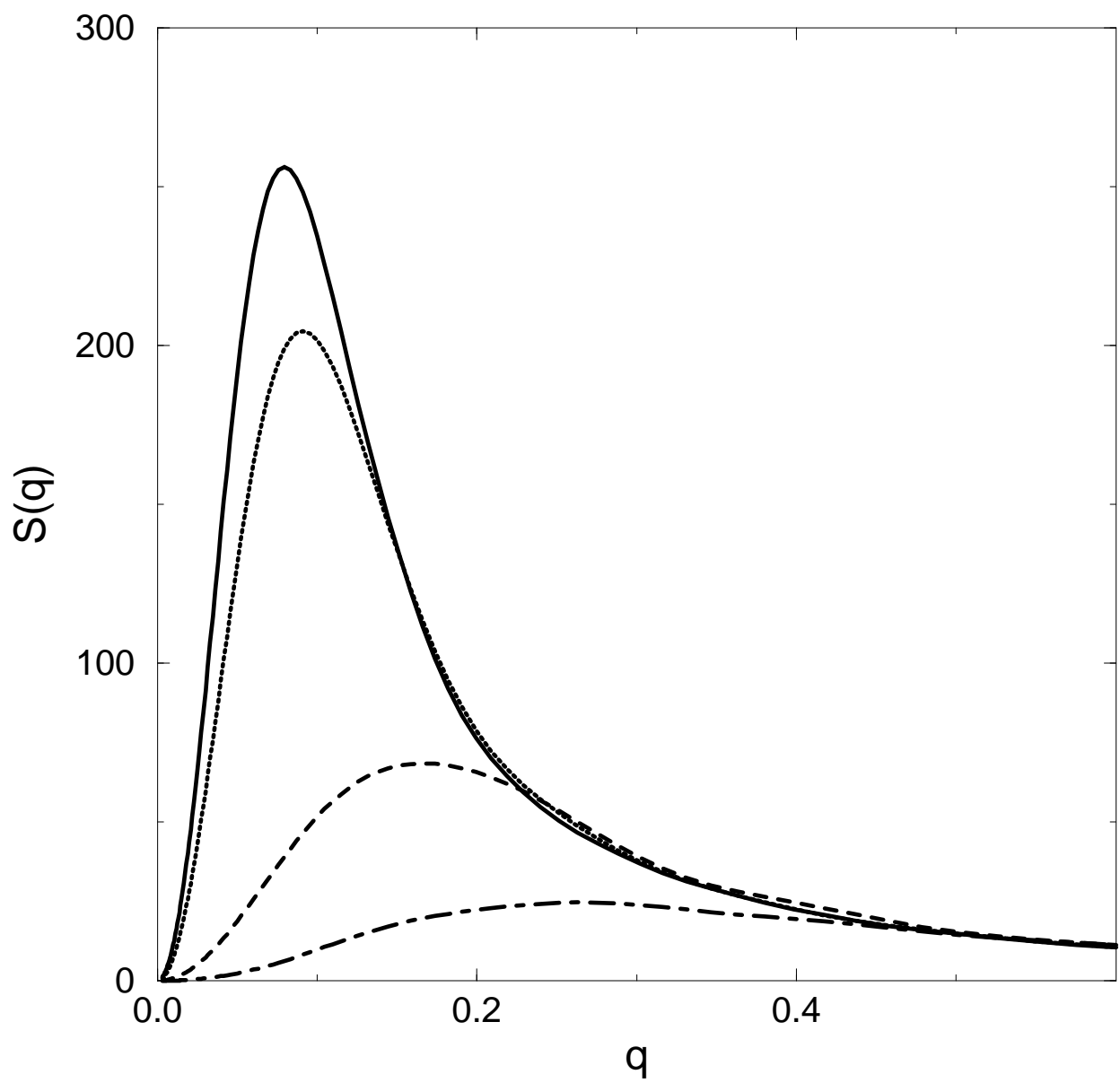
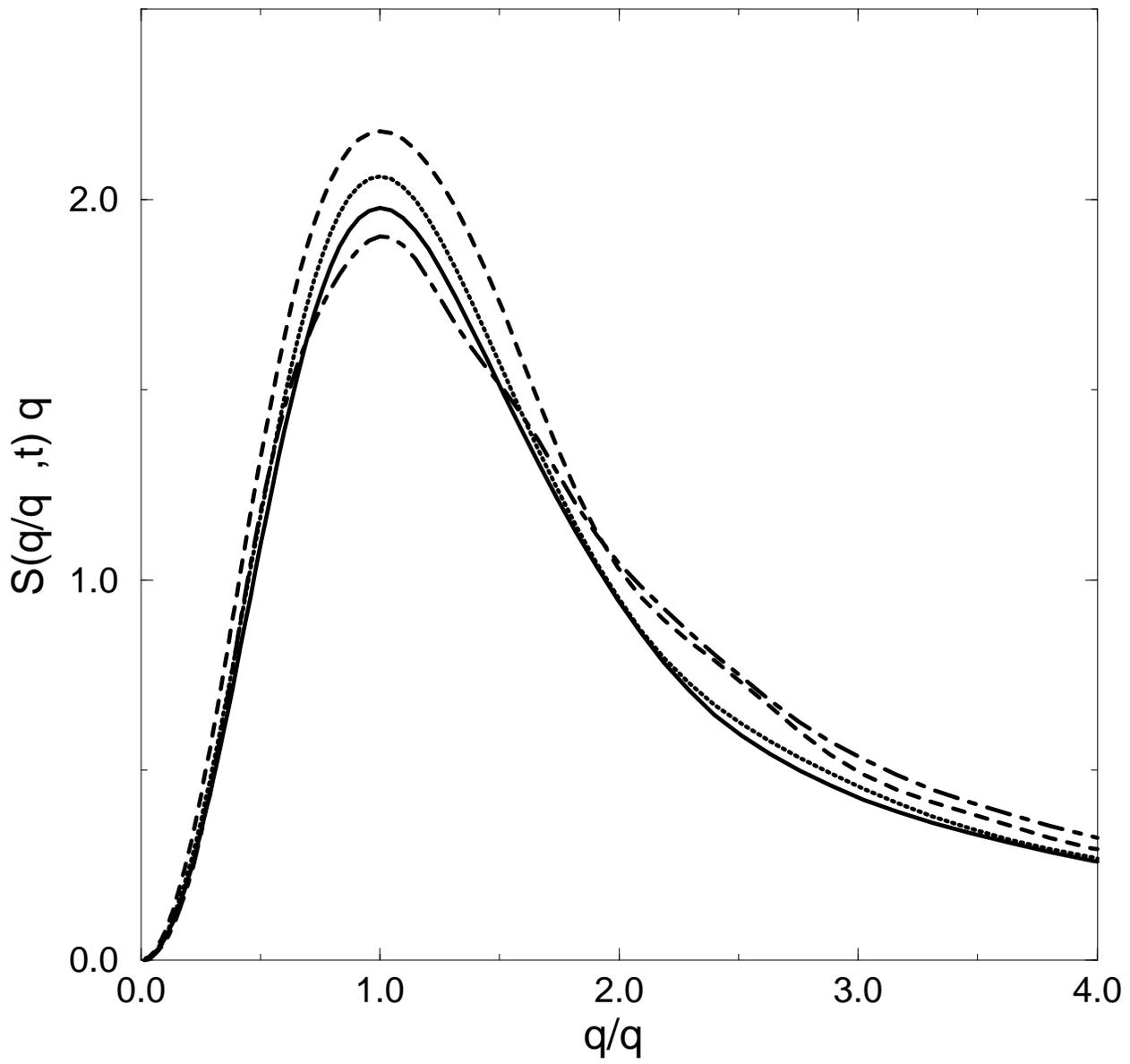
\end{document}